\documentclass[%
  aps,%
  prl,%
  superscriptaddress,%
  reprint,%
  notitlepage]{revtex4-1}
\relax

\usepackage{amsmath}
\usepackage{amssymb}
\usepackage{amsthm}
\usepackage{graphicx}
\usepackage{mathtools}
\usepackage{enumitem}
\usepackage{microtype}
\usepackage{bbm}
\usepackage[autostyle,autopunct=true]{csquotes}

\usepackage{xcolor}

\usepackage{hyperref}
\usepackage[capitalize,nameinlink]{cleveref}

\definecolor{docnotelinkcolor}{rgb}{0,0,0.4}%
\hypersetup{%
  bookmarksnumbered=false,bookmarksopen=false,bookmarksopenlevel=1,%
  breaklinks=true,pdfborder={0 0 0},colorlinks=true,%
  anchorcolor=docnotelinkcolor,citecolor=docnotelinkcolor,%
  filecolor=docnotelinkcolor,linkcolor=docnotelinkcolor,%
  menucolor=docnotelinkcolor,runcolor=docnotelinkcolor,%
  urlcolor=docnotelinkcolor}

\newtheoremstyle{prlthm}%
  {0pt}%
  {0pt}%
  {\normalfont}%
  {\parindent}%
  {\itshape}%
  {:}%
  { }%
  {\thmname{#1\thmnumber{ #2}\thmnote{ #3}}}%

\theoremstyle{prlthm}
\newcounter{thm}

\DeclareMathOperator{\tr}{tr}

\DeclarePairedDelimiter\norm{\lVert}{\rVert}
\DeclarePairedDelimiter\avg{\langle}{\rangle}

\newcommand\iid{i.i.d.\@}

\begin{document}

\title{Thermodynamic Capacity of Quantum Processes}

\author{Philippe Faist}
\email{phfaist@caltech.edu}
\affiliation{Institute for Quantum Information and Matter, Caltech, Pasadena, CA 91125, U.S.A.}

\author{Mario Berta}
\affiliation{Department of Computing, Imperial College London, London SW7 2AZ, United Kingdom}

\author{Fernando Brand\~ao}
\affiliation{Google Inc., Venice, CA 90291, U.S.A.}
\affiliation{Institute for Quantum Information and Matter, Caltech, Pasadena, CA 91125, U.S.A.}

\date{May~24, 2019}

\begin{abstract}
  Thermodynamics imposes restrictions on what state transformations are
  possible. In the macroscopic limit of asymptotically many independent copies
  of a state---as for instance in the case of an ideal gas---the possible
  transformations become reversible and are fully characterized by the free
  energy.  In this Letter, we present a thermodynamic resource theory for
  quantum processes that also becomes reversible in the macroscopic limit, a
  property that is especially rare for a resource theory of quantum channels.
  We identify a unique single-letter and additive quantity, the thermodynamic
  capacity, that characterizes the ``thermodynamic value'' of a quantum channel,
  in the sense that the work required to simulate many repetitions of a quantum
  process employing many repetitions of another quantum process becomes equal to
  the difference of the respective thermodynamic capacities.  On a technical
  level, we provide asymptotically optimal constructions of universal
  implementations of quantum processes.  A challenging aspect of this
  construction is the apparent necessity to coherently combine thermal
  engines that would run in different thermodynamic regimes depending on the
  input state.
  Our results have applications in quantum Shannon theory by providing a
  generalized notion of quantum typical subspaces and by giving an operational
  interpretation to the entropy difference of a
  channel.  %
\end{abstract}

\maketitle

\begin{figure}
  \centering
  \includegraphics{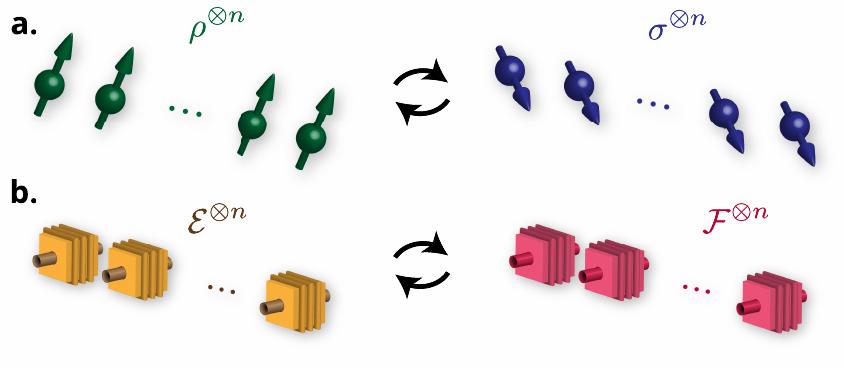}
  \vspace*{-5mm} %
  \caption{Reversibility in the many-copy regime.
    \textbf{a.}~In the resource theory of thermodynamics, quantum states are
    reversibly interconvertible, i.e., the work cost of transforming $n$ independent
    copies of $\rho$ into $n$ independent copies of $\sigma$ is (approximately) the
    same as the work that can be extracted in the reverse transformation.
    Reversibility is a valuable property for a resource theory, as it provides a
    full characterization of the precise amount of resources that is necessary
    for any state transformation~\protect\cite{Chitambar2019RMP_resource}.
    \textbf{b.}~We show that a similar conclusion holds for quantum
    processes. There is a unique quantity, the thermodynamic capacity, that
    measures the ``thermodynamic value'' of the channel in terms of the
    resources required to create, or extracted while consuming, many copies of a
    channel.  Note that, in this context, reversibility refers to the
    interconversion of processes themselves, not to recovering the input of a
    process from its output.}
\label{fig:AsymptReversibleInterconversion}
\end{figure}

\emph{Introduction}.---%
In the quest of extending the laws of thermodynamics beyond the macroscopic
regime, the resource theory of thermal operations was introduced to characterize
possible transformations which could be carried out at the nano
scale~\cite{Brandao2013_resource,Horodecki2013_ThermoMaj,Brandao2015PNAS_secondlaws,Gour2018NatComm_entropic,Chubb2018Qu_beyond}. By imposing a set of physically motivated rules, an agent can only
perform a restricted set of operations on a system, which we refer to
generically as \emph{thermodynamic operations}.  Here, we will consider as
thermodynamic operations either thermal operations~\cite{Brandao2013_resource,Horodecki2013_ThermoMaj,Brandao2015PNAS_secondlaws} or Gibbs-preserving
maps~\cite{Janzing2000_cost,Faist2015NJP_Gibbs,Faist2015NatComm,Faist2018PRX_workcost}.  By characterizing the possible state
transformations under these rules, one obtains formulations of the second law of
thermodynamics which are valid for small-scale systems out of thermodynamic
equilibrium. A natural regime to study such state transformations is a
macroscopic regime in which one considers conversions between many independent
and identically distributed (i.i.d.\@) copies of a state, i.e., states of the
form $\rho^{\otimes n}$. If we consider transformations on a system $S$ with
Hamiltonian $H_S$, using a heat bath at inverse temperature $\beta$ and a work
storage system $W$, then the asymptotic work cost per copy of transforming
$\rho^{\otimes n}$ into $\sigma^{\otimes n}$ is given by the difference in free
energy $F(\sigma) - F(\rho)$~\cite{Brandao2013_resource}. The free energy is
defined as
\begin{align}
  F(\rho)
  = \tr(H \rho) - \beta^{-1} S(\rho)
  = \beta^{-1} D( {\rho} \| {e^{-\beta H}} )\,,
\end{align}
expressed either in terms the von Neumann entropy $S(\rho) = -\tr(\rho\ln\rho)$
or the quantum relative entropy
$D(\rho \| \gamma) = \tr[\rho(\ln\rho -\ln\gamma)]$. Since the cost of
asymptotically performing the reverse transformation
$\sigma^{\otimes n} \to \rho^{\otimes n}$ is the negative of the cost of the
forward transformation this resource theory becomes reversible
(\cref{fig:AsymptReversibleInterconversion}\textbf{a}).

Here, we study the resource theory of thermodynamics for quantum processes
themselves. Given a black-box implementation of a process $\mathcal{E}$, can we
simulate a process $\mathcal{F}$ using thermodynamic operations, or is there a
thermodynamic cost in doing so? We fully answer the question in the i.i.d.\@
regime, and show that the thermodynamic simulation of channels becomes
reversible
(\cref{fig:AsymptReversibleInterconversion}\textbf{b}). That is, the
work cost of executing many realizations of $\mathcal{F}$ using many
realizations of $\mathcal{E}$ is the same as the work that can be extracted in
the reverse process of implementing $\mathcal{E}$ from uses of $\mathcal{F}$.

Reversibility is a coveted property for a resource
theory~\cite{Chitambar2019RMP_resource}, as it usually follows from this
property that one can establish necessary and sufficient conditions for the
interconversion of states, thus fully understanding which state transformations
are possible.  On the other hand, many natural resource theories do not have
this property.  For instance, entanglement transformations under local
operations and classical communication is a reversible resource theory only for
the set of pure states, and not for general mixed
states~\cite{Bennett1996PRA_concentrating,Horodecki1998PRL_bound}.  Also,
resource theories of channels are typically not asymptotically
reversible~\cite{Liu2019arXiv_channels} except in certain specific cases, for
instance simulating a quantum channel with local operations, classical
communication and with arbitrary shared
entanglement~\cite{Bennett2002IEEETIT_reverse,Berta2011_reverse,Bennett2014_reverse}.

\begin{figure}
  \centering
  \includegraphics{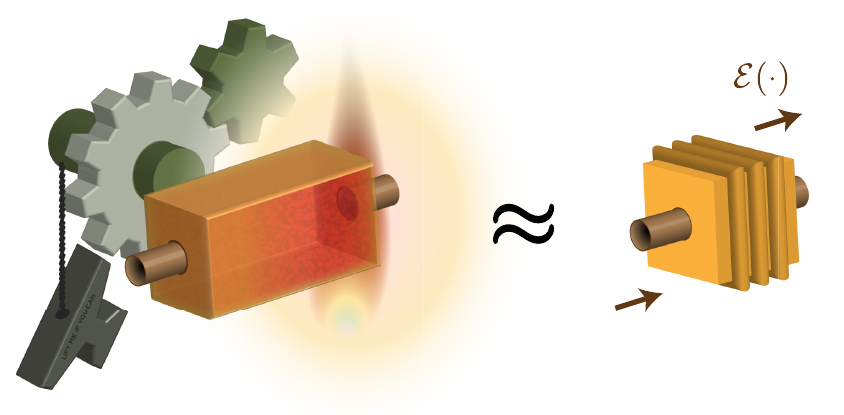}
  \caption{Universal thermodynamic implementation of a quantum process. Using
    thermodynamic operations and by furnishing work, the task is to simulate
    approximately an ideal process $\mathcal{E}$.  The implementation must
    output a state close to $\mathcal{E}(\sigma)$ for any possible input
    $\sigma$, even relative to a reference system.  In this letter, we consider
    the regime of many independent copies of the channel, the i.i.d.\@ regime,
    and we show that it is possible to construct a universal implementation of
    an \iid{} quantum process at a work cost rate equal to the
    \emph{thermodynamic capacity} of the channel, defined as the worst-case
    difference of free energies between the output and the input.  The main
    challenge is that the implementation must work for any input state,
    including superpositions of input states for which individual input-specific
    implementations would run in different thermodynamic regimes.}
  \label{fig:SimulationOfChannel}
\end{figure}

\emph{Challenges}.---%
The main problem we would like to solve is to find a universal implementation of
any quantum process in the \iid{} regime using thermodynamic operations
(\cref{fig:SimulationOfChannel}).
In the single-shot setting, where the process is only performed once and the
implementation must succeed with high probability, we can leverage results
from~\cite{Faist2018PRX_workcost} to characterize the corresponding work cost
for any input state.  The single-shot work cost for a known input state $\sigma$
is given by the \emph{coherent relative entropy}~\cite{Faist2018PRX_workcost},
which in the \iid{} limit is known to converge to the free energy difference
$F(\mathcal{E}(\sigma)) - F(\sigma)$.  If we adapt the definition of the
coherent relative entropy by imposing that the implementation is accurate for
any input state, we obtain a mathematical expression that characterizes the work
cost of a single-instance universal implementation of a process.
However, it is unclear what this cost becomes in the \iid{} limit.  More
precisely, it is not clear that one can find an implementation of an \iid{}
process that performs accurately for any input state, because the implementation
seems to have to be able to work coherently in a superposition of different
thermodynamic regimes of energy-to-entropy conversion.
To see this, consider the following naive attempt at an implementation of an
\iid{} process $\mathcal{E}^{\otimes n}$.  One could determine the input state
nondestructively using gentle tomography~\cite{Bennett2006PRA_nondestructive},
and then apply the implementation $\mathcal{T}_{\sigma}$ given
by~\cite{Faist2018PRX_workcost} that is optimal for the given \iid{} input state
$\sigma^{\otimes n}$.  Mathematically, this is:
\begin{align}
  \mathcal{T}(\cdot)
  = \sum_{\sigma} \mathcal{T}_\sigma( {M}_{\sigma} (\cdot) {M}_{\sigma}  )\ ,
\end{align}
where ${M}_{\sigma}$ would be the square root of the positive operator-valued measure
(POVM) effects corresponding
to the gentle tomography of~\cite{Bennett2006PRA_nondestructive}, which
estimates approximately which \iid{} state $\sigma^{\otimes n}$ the input is
while not disturbing the state too much.  Crucially, the different optimal
processes $\mathcal{T}_\sigma$ might have different work costs, and worse, might
be working in different regimes of energy-to-entropy conversion.  For instance,
the ideal process $\mathcal{E}$ might be a combination of a Landauer erasure for
one input but a shift in energy levels for a different input.  So what happens
if a superposition of different \iid{} inputs is given?  In this case, the state
is decohered by the gentle tomography step, destroying the superposition.  So we
need to somehow find a way of combining the different $\mathcal{T}_\sigma$
coherently, with the difficulty that each of the processes $\mathcal{T}_\sigma$
for different $\sigma$ might be operating in different thermodynamic regimes.

At the heart of this problem is the fact that we require the implementation to
be accurate even for non-\iid{} inputs.  For instance, if we are developing a
specialized computing chip that implements a given quantum gate in a quantum
computer, the different inputs to the gate might very well be correlated, as
they would depend on earlier stages of the computation.
Here, we may appeal to the postselection
technique~\cite{Christandl2009PRL_Postselection}, which essentially states that
if an implementation $\mathcal{T}_n$ of an \iid{} process
$\mathcal{E}^{\otimes n}$ is accurate for any \iid{} input state with an
accuracy that decreases exponentially in the number of copies $n$, then the
implementation $\mathcal{T}_n$ is also accurate for any input state and not only
for \iid{} inputs.  %
Unfortunately, this actually even precludes the use of nondestructive or gentle
tomography as suggested above, because the error induced by this step is
polynomial in $n$ instead of exponential.
So we are back to square one.

A different natural approach is to follow the proof logic of
Refs.~\cite{Berta2011_reverse,Berta2014IEEETIT_InfGainMeas}: One can define an
entropic quantity associated with the single-shot problem instance, as described
above, and exploit properties of the relevant entropy measures.  In fact, this
proof strategy works for systems described by a trivial Hamiltonian
$H=0$. There, the single-shot work cost for i.i.d.\@ processes reduces to a
single conditional max-entropy quantity using the post-selection
technique~\cite{Christandl2009PRL_Postselection}.  Then, exploiting a
quasi-convexity-like property of the conditional
max-entropy~\cite{Morgan2014IEEETIT_prettystrong} allows to prove our result for
the special case of trivial Hamiltonians (technical details will be published
elsewhere~\cite{TechnicalPaperThCapacity}).  Attempting to generalize this proof
approach to non-trivial Hamiltonians fails because the coherent relative entropy
does not display the required quasi-convexity property.

\emph{Main result}.---%
We solve the above problems by explicitly constructing a universal
implementation of any \iid{} quantum process, using a new notion of quantum
typicality that intuitively allows to coherently combine different
entropy-to-energy conversion regimes.  By construction, the implementation does
not depend on the input state, and when considered as a channel, it is close in
diamond norm to the ideal channel $\mathcal{E}^{\otimes n}$~\footnote{The
  diamond norm is defined for two channels
  \protecting{$\mathcal{E},\mathcal{F}$} acting on the same system as
  \protecting{$\norm{\mathcal{E}-\mathcal{F}}_\diamond =
    \max_\sigma\,\protect\norm[\big]{ (\mathcal{E}\otimes\mathrm{id})(\sigma) -
      (\mathcal{F}\otimes\mathrm{id})(\sigma)}_1$}, where
  \protecting{$\norm{X}_1:= \tr\bigl(\sqrt{X^\dagger X}\bigr)$} denotes the
  trace norm.}.  The rate at which work has to be supplied is characterized by
the \emph{thermodynamic capacity} of the channel, given as
\begin{align}
  \label{eq:capacity}
  T(\mathcal{E})
  = \max_\sigma \, \Bigl\{ F(\mathcal{E}(\sigma)) - F(\sigma) \Bigr\}\,.
\end{align}
That is, the work cost of such an implementation coincides with the worst-case
cost of implementing the process over all possible i.i.d.\@ input states.
Surprisingly in light of the above challenges, there is no intrinsic overhead in
the work cost rate associated with the implementation being ignorant of the
input state---the work cost rate is no worse than the rate corresponding to the
worst-case input state.

We may combine our main result with the following result by Navascu\'es \emph{et
  al.}~\cite{Navascues2015PRL_nonthermal}: From a black-box access to many
copies of a process $\mathcal{E}$, it is possible to extract work at a rate that
is asymptotically equal to $T(\mathcal{E})$.  Hence the work cost rate
associated with converting $\mathcal{E}^{\otimes n}$ to
$\mathcal{F}^{\otimes n}$ is simply $T(\mathcal{F}) - T(\mathcal{E})$,
corresponding to extracting as much work as possible from the copies of
$\mathcal{E}$ and then simulating $\mathcal{F}^{\otimes n}$ using our procedure.
This work cost is reversible, i.e., all the work invested in the transformation
can be recovered in the reverse transformation
$\mathcal{F}^{\otimes n}\to\mathcal{E}^{\otimes n}$.  Hence, this work cost is
optimal, and the resource theory becomes reversible, with the thermodynamic
capacity being the unique measure of the ``thermodynamic value'' of the quantum
channel (see also \cref{fig:AsymptReversibleInterconversion}).

The thermodynamic capacity~\eqref{eq:capacity} generalizes the notion of
capacity for quantum channels to the context of thermodynamics, by measuring how
much free energy can be conveyed through the use of the channel.  The
thermodynamic capacity is expressed as a single-letter formula and can be
computed efficiently, even analytically for some simple examples, as it can be
formulated as a convex optimization
problem~\cite{BookBoyd2004ConvexOptimization}.  The thermodynamic capacity is
also additive~\cite{Navascues2015PRL_nonthermal}, and does not need to be
regularized as for other channel capacity measures.
It is tightly related to other channel entropy measures, especially the
amortized entanglement of a channel~\cite{Kaur2018JPA_amortized} and the entropy
of a channel~\cite{Gour2018arXiv_entropychannel}.

We announce our results in this Letter, providing the physical background and
intuition surrounding our construction of a universal implementation of any
\iid{} process.  Fully detailed technical proofs will be published
elsewhere~\cite{TechnicalPaperThCapacity}.

\emph{Implementation based on quantum typicality}.---%
We exploit two main ingredients for our universal implementation.  First, using
Schur-Weyl duality one may estimate approximately the spectrum of an \iid{}
state, and hence its entropy, using a global measurement on the $n$
systems~\cite{PhDHarrow2005,Bennett2006PRA_nondestructive,Bennett2014_reverse,Haah2017IEEETIT_sampleoptimal}.  Let us denote by $\{ P_{s}^n \}_s$ the
POVM that produces an estimate $s$ of the
entropy per copy of an \iid{} state over $n$ systems.  Second, for
noninteracting systems, a global energy measurement will provide a sharp
statistics for any \iid{} state due to large deviation bounds (cf.\@ for
example~\cite{YungerHalpern2016NatComm_NATSandNATO}); thus the average energy
per copy $h$ of an \iid{} state can be estimated to a good approximation by this
measurement, whose POVM effects we denote by $\{ Q_{h}^n \}$.

Our construction is then based on the following idea.  Given an \iid{} process
$\mathcal{E}_{X\to X'}$, we may consider a Stinespring isometry $V_{X\to X'E}$
into an environment $E$, satisfying
$\mathcal{E}(\cdot) = \tr_E\bigl(V (\cdot) V^\dagger)$.  Using the POVMs
mentioned above, we may write
$V = (\sum_{s',h'} Q_{h'} P_{s'}) V (\sum_{s,h} P_{s} Q_{h})$, since the
elements of a POVM sum to the identity operator.  An important observation
however is that not all combinations of outcomes $s,h,s',h'$ are possible.
Namely, we know that for any \iid{} input state $\sigma^{\otimes n}$, we must
have
$-s' + \beta h' + s - \beta h \approx -S(\mathcal{E}(\sigma)) + \beta
\avg{H_{X'}}_{\mathcal{E}(\sigma)} + S(\sigma) - \beta \avg{H_{X}}_{\sigma} =
\beta F(\mathcal{E}(\sigma)) - \beta F(\sigma) \leqslant \beta
T(\mathcal{E})$.  So we may enforce this condition explicitly in the
decomposition above, and by pushing all the POVM effects through the isometry,
we get a candidate implementation of the form
$\mathcal{T}_{X^n\to X'^n} = \tr_{E^n}\bigl( W\,(\cdot)\,W^\dagger)$ with
$W_{X\to X'E} = M_{X'E}\,V_{X\to X'E}$, and where $M$ is defined as
\begin{align}
  M_{X'E} = \sum_{\substack{-s'+\beta h'+s-\beta h\lesssim \beta T(\mathcal{E})}}
  Q_{h'} P_{s'} \bigl(V P_{s} Q_{h} V^\dagger\bigr)\ .
  \label{eq:our-typical-operator}
\end{align}
The operator $M$ can be interpreted as a fully quantum, universal smoothing
operator for bipartite states that counts entropy relative to another operator,
and which is a natural generalization of universal and relative typical subspace
projectors~\cite{Jozsa1998PRL81,Hayashi2002PRA_variable,Jozsa2003,Bennett2006PRA_nondestructive,Bjelakovic2003arXiv_revisted,Noetzel2012arXiv_two,BookWilde2013QIT,Bennett2014_reverse,Berta2015QIC_monotonicity,Sen2018arXiv_joint}.  The approximations above are
related to how well the POVMs $\{ P_s^n \}$ and $\{ Q_h^n \}$ are able to
resolve the entropy and energy per copy, and the corresponding error vanishes in
the limit $n\to\infty$.
That the implementation process $\mathcal{T}_{X^n\to X'^n}$ is close in diamond
norm to the ideal \iid{} process $\mathcal{E}^{\otimes n}$ follows from the
postselection technique~\cite{Christandl2009PRL_Postselection}, which allows us
to focus on \iid{} input states, and from the fact that the omitted terms
in~\eqref{eq:our-typical-operator} only account for an exponentially small
weight for any input \iid{} state.  Finally, we invoke a mathematical
characterization of the work cost of processes in the framework of
Gibbs-preserving maps~\cite{Faist2018PRX_workcost}: We show that
$\mathcal{T}(e^{-\beta H_{X^n}}) \lesssim
e^{nT(\mathcal{E})}e^{-\beta{}H_{X'^n}}$, which in turn implies that the
candidate implementation $\mathcal{T}$ requires $T(\mathcal{E})$ work per copy.
The complete proof of these statements is provided in
Ref.~\cite{TechnicalPaperThCapacity}.

\emph{Thermal Operations.}---%
The framework of Gibbs-sub-preserving maps is particularly generous, and it is a
priori not clear that all such maps are implementable at no work cost.
Alternatively, we may consider the framework of thermal operations, where only
energy-conserving unitary interactions with a heat bath are allowed.  It turns
out that in this framework it is also possible to construct a universal
implementation of \iid{} processes at a work cost of $T(\mathcal{E})$ per copy,
yet our proof is restricted to processes that are time-covariant, i.e., that
commute with the time evolution.  The assumption of time covariance allows us
to sidestep issues of coherence between energy
levels~\cite{Marvian2014PRA_asymmetry,Aberg2014PRL_catalytic,Lostaglio2015PRX_coherence,Cirstoiu2017arXiv_gauge,Marvian2018arXiv_distillation}. %
This result directly implies that the asymptotic reversibility of the resource
theory of quantum processes also holds in the context of thermal operations for
time-covariant processes.
Our proof for this second main result is presented in detail in
Ref.~\cite{TechnicalPaperThCapacity}, and follows a considerably different
strategy than above; we make use of recent ideas from quantum information
theory, including the convex-split lemma and position-based
decoding~\cite{Anshu2017PRL_convexsplit,Anshu2019IEEETIT_oneshot,Anshu2018IEEETIT_redistribution,Anshu2018IEEETIT_SlepianWolf,Anshu2019IEEETIT_compression,Majenz2017PRL_catdecoupling}.

\emph{Extensions.}---%
Our proof techniques allow us to prove some related results.
First, we exhibit a one-shot conditional erasure protocol that is valid for
systems described by a non-trivial Hamiltonian and states that are
time-covariant, thus generalizing the protocol of Ref.~\cite{delRio2011Nature}.
The work cost is given in terms of a beyond-i.i.d.\@ generalization of the
relative entropy called the \emph{hypothesis testing relative
  entropy}~\cite{Wang2012PRL_oneshot,Tomamichel2013_hierarchy,Matthews2014IEEETIT_blocklength,Buscemi2010IEEETIT_capacity,Brandao2011IEEETIT_oneshot,Dupuis2013_DH}, that quantifies how well two states can be distinguished by a
hypothesis test and which is closely-related to other one-shot information
measures~\cite{PhDRenner2005_SQKD,Datta2009IEEE_minmax,PhDTomamichel2012,BookTomamichel2016_Finite}.  Our result
implies that it is possible to implement any time-covariant process for a fixed
time-covariant input state in the single-shot regime, using thermal operations
and a battery, at a cost given by the coherent relative entropy.

Also, we show that if the input is a fixed i.i.d.\@ state, it is possible to
implement any arbitrary, not necessarily time-covariant i.i.d.\@ channel using
thermal operations, a battery, and a sub-linear amount of coherence, at the same
asymptotic work cost per copy as it would take to implement it with
Gibbs-preserving maps.  We thus conclude that although Gibbs-preserving maps are
more powerful in general than thermal operations~\cite{Faist2015NJP_Gibbs}, they
become asymptotically equivalent in the macroscopic limit in terms of
implementing i.i.d.\@ processes on given i.i.d.\@ input states.

\emph{Discussion.}---%
Quantum resource theories of channels has developed into a hot topic of interest
in recent years, as they display various features that are not mirrored in
corresponding resource theories of state transformations~\cite{Fang2019IEEETIT_simulation,Anshu2018arXiv_partially,Liu2019arXiv_channels,Gour2018arXiv_entropychannel}.  In this context, we show
that the thermodynamic resource theory of channels is asymptotically reversible
in the \iid{} regime, as is the case for quantum states.
Asymptotically, there exists a unique monotone, the thermodynamic capacity,
which characterizes the ``thermodynamic value'' of a channel.  In this sense our
result is the thermodynamic analogue of the reverse Shannon theorem for quantum
channels~\cite{Bennett2014_reverse,Berta2011_reverse}.

Our statements and proofs are also fully noncommutative in the sense that they
cannot be simplified to a problem about classical probability distributions in a
fixed basis---a feature that is still rather uncommon in quantum thermodynamics.
Our universal implementation of an \iid{} process is accurate even for
superpositions of input states for which individual optimal implementations
would require thermal engines running in different regimes of energy-to-entropy
conversion.
Moreover, standard proof techniques developed for quantum channel simulations do
not readily apply to our problem at hand; attempting to mimick the proof in
Refs.~\cite{Berta2011_reverse,Berta2014IEEETIT_InfGainMeas} fails because the
coherent relative entropy is not quasi-convex~\cite{TechnicalPaperThCapacity}.

Whether or not it is possible to universally implement any i.i.d.\@ channel that
is not time-covariant using thermal operations is still an open question. We
expect that such a protocol might in general need a very large amount of
coherence, much like the requirement of large embezzling states for the reverse
Shannon theorem~\cite{Bennett2014_reverse,Berta2011_reverse}. Indeed, if the
input is a superposition of two different i.i.d.\@ states of different energy,
the environment must be able to coherently compensate for any energy difference
caused by the process without disturbing the process. 
However, we have shown that for fixed \iid{} input states, any i.i.d.\@ channel
can be implemented optimally using thermal operations, so this suggests that a
tighter connection between thermal operations and Gibbs-preserving maps remains
to be uncovered.

For a trivial Hamiltonian, the thermodynamic capacity reduces (up to a sign) to
the \emph{minimal entropy gain} of a channel $\mathcal{E}$, defined as
$\min_\sigma[S(\mathcal{E}(\sigma)) - S(\sigma)]$.  This quantity was introduced
as a measure of information for channels~\cite{Alicki2004arXiv_isotropic,Devetak2006CMP_multiplicativity,Holevo2011ISIT_entropygain,Holevo2010DM_infinitedim,Holevo2011TMP_CJ,BookHolevo2012_QuSystemsChannelsInformation,Buscemi2016PRA_reversibility}.  Our results thus exhibit a physical and
operational interpretation for this quantity.

Given the relevance of %
entropy measures for a wide range of physical and information-theoretic
situations, we expect our results to find applications beyond thermodynamic
interconversion of processes. For instance, we note that a quantity closely
related to the coherent relative entropy has found applications in studying
dissipative dynamics of many-body
systems~\cite{Capel2018JPA_conditional}. Also, in contrast to standard smooth
entropy measures for quantum states~\cite{PhDTomamichel2012}, our channel
smoothing in terms of the diamond norm leaves one of the marginals invariant
when applied to quantum states (cf.~the very recent related
works~\cite{Fang2019IEEETIT_simulation,Anshu2018arXiv_partially}). This might offer some insights on the quantum
joint typicality conjecture in quantum communication
theory~\cite{PhDDutil2011_multiparty,Noetzel2012arXiv_two,Drescher2013ISIT_simultaneous}, on which recent process has been
made~\cite{Sen2018arXiv_joint}.
One may also study how our results are modified if we replace the diamond norm
condition on the implementation by other channel distinguishability measures,
such as introduced in Ref.~\cite{Chiribella2008PRL_memory}; this would be
particularly relevant for settings with memory effects, for instance
implementations of gates in a quantum computer.

Finally, that there exists optimal universal thermodynamic implementations of
channels indicates that low-dissipation components for future quantum devices
can in principle be developed, that function accurately for all inputs, and
still dissipate no more than required by the worst case input.

\emph{Acknowledgments.}---%
We thank \'Alvaro Alhambra, David Ding, Patrick Hayden, Rahul Jain, David
Jennings, Mart\'\i{} Perarnau-Llobet, Mark Wilde, and Andreas Winter for
discussions.  PhF acknowledges support from the Swiss National Science
Foundation (SNSF) through the Early PostDoc.Mobility Fellowship No.\@
{P2EZP2\_165239} hosted by the Institute for Quantum Information and Matter (IQIM)
at Caltech, from the IQIM which is a National Science Foundation (NSF) Physics
Frontiers Center (NSF Grant PHY-{1733907}), and from the Department of Energy
Award {DE}-{SC0018407}.
FB is supported by the NSF.

\def\ {\unskip\space}
\def\doibase#110.{https://doi.org/10.}

\end{document}